\documentstyle[11pt,aaspp4,flushrt]{article}
\catcode`\@=11 
\def\@versim#1#2{\vcenter{\offinterlineskip
        \ialign{$\m@th#1\hfil##\hfil$\crcr#2\crcr\sim\crcr } }}
\newcommand{\beq}{\begin{equation}}
\newcommand{\eeq}{\end{equation}}

\def\lsim{\mathrel{\mathpalette\@versim<}}
\def\gsim{\mathrel{\mathpalette\@versim>}}

\begin{document}
\title{Chandra, GLAST, and the Galactic Center}
\author{Eliot Quataert\footnote{Chandra Fellow} and Andrei Gruzinov}
\affil{Institute for Advanced Study, School of Natural Sciences, Einstein Drive, Princeton, NJ 08540; eliot@ias.edu, andrei@ias.edu}

\begin{abstract}
Two-temperature spherical accretion flows produce $\approx 100$ Mev
gamma-rays from the decay of neutral pions created in proton-proton
collisions close to the black hole; they also produce $\sim 10$ keV
X-rays by bremsstrahlung emission at large radii.  The gamma-ray to
X-ray luminosity ratio is nearly independent of black hole mass and
accretion rate.  It does depend sensitively on the radial density
profile of the accretion flow through the parameter $a$, where $n
\propto r^{-a}$.  For the canonical Bondi value of $a = 3/2$, the
gamma-ray to X-ray luminosity ratio is $\approx 30$.  We interpret a
recent Chandra detection coincident with the massive black hole at the
Galactic Center as being thermal bremsstrahlung emission from the
accretion flow.  With this normalization, the expected gamma-ray
luminosity is $\approx 10^{35}$ ergs s$^{-1}$ if $a = 3/2$.  This is
nearly two orders of magnitude above the detection threshold of the
GLAST telescope.  For $a \approx 1/2$, however, (a value suggested by
recent theoretical arguments), the expected gamma-ray luminosity is
only $\approx 10^{29}$ ergs s$^{-1}$; GLAST should therefore provide
an important probe of the true accretion rate and radial density
profile of the accretion flow onto Sgr A*.

\

\noindent {\it Subject Headings:} accretion, accretion disks ---
Galaxy: center --- gamma rays: theory

\end{abstract} 

\section{Introduction}

In roughly spherical accretion flows, be it Bondi (1952) or
advection-dominated accretion flows (ADAFs; Rees et al. 1982, Narayan
\& Yi 1994), the protons have temperatures comparable to their
gravitational potential energy.  Close to the black hole, a
significant number of protons are energetic enough to exceed the
threshold for the production of pions in proton-proton collisions.
Neutral pions quickly decay to produce gamma-rays.  It has long been
recognized that this is a plausible source of gamma-ray emission from
spherical accretion flows (Shvartsman 1971; Dahlbacka, Chapline, \&
Weaver 1974; Colpi, Maraschi, \& Treves 1986; Mahadevan, Narayan, \&
Krolik 1997).

In this paper we place the expected gamma-ray emission from spherical
accretion flows on firmer observational ground by relating it to the
more readily observable x-ray emission produced by thermal
bremsstrahlung.  A simple and relatively universal relationship
between the two fluxes exists because both are produced by two-body
processes.  This is discussed in the next section (\S2).  We then
apply these considerations to Sgr A*, the supermassive black hole at
the center of our galaxy (\S3).  In \S4 we briefly summarize our
results.

\section{The Gamma-ray to X-ray Luminosity Ratio}

We first give a simple calculation of the gamma-ray to x-ray
luminosity ratio assuming self-similar scalings for the density and
temperature of the flow and a thermal distribution of protons.  We
then discuss the uncertainties introduced by these approximations.  

We take the temperature and number density of the flow to be \beq
\theta_p = \theta_0 \ r^{-1} \ \ \ {\rm and} \ \ \ n = n_0 \ r^{-a},
\label{td} \eeq where $\theta_p = kT_p/m_p c^2$ is the dimensionless
proton temperature, $r$ is the radius in the flow in units of the
Schwarzschild radius ($R_S$), and $n_0$ is the normalization of the
density, which depends on, e.g., the black hole mass, the accretion
rate, and the viscosity parameter $\alpha$.  For $\theta_0 = 0.15$,
the proton temperature profile is that of non-relativistic Bondi
accretion with an adiabatic index of $\gamma =
5/3$;\footnote{Relativistic corrections are small.} comparable maximal
temperatures and identical radial scalings occur in relativistic Bondi
accretion (Shapiro 1973) and ADAFs (Narayan \& Yi 1995).  The electron
temperature profile is rather uncertain; fortunately we will only need
the electron temperature at large radii, $r \gsim 10^3$, where the
flow is well approximated as one temperature.  In equation (\ref{td})
we allow the radial density profile to differ from the canonical Bondi
value of $a = 3/2$; recent work on ADAFs has shown that much smaller
values, e.g., $a = 1/2$, may be appropriate (see \S2.2).

The number of $\approx 100$ Mev gamma-rays produced per second and per
cm$^3$ is given by $n^2 R(\theta_p)$, where $R(\theta_p)$ is the
reaction coefficient for thermal protons of temperature $\theta_p$.
At $\theta_p \approx 0.15$, $R(\theta_p)$ can be approximated by
$R(\theta_p) \approx R_0 (\theta_p/0.15)^3$, where $R_0 \approx 2
\times 10^{-17}$ cm$^3$ s$^{-1}$ (see Fig. 3 of Dermer 1986); for
$\theta_p \lsim 0.05$, $R(\theta_p)$ decreases much more rapidly than
$\propto \theta_p^3$.  Integrating over the flow, the photon
luminosity in $\approx 100$ Mev gamma-rays is \beq N_\gamma = 4 \pi
R_S^3 n_0^2 \int_{1}^{\infty} \left({dr \over r}\right) R(\theta)
r^{3-2a} \approx {2 \pi \over a} R^3_S n^2_0 R_0
\label{ngamma}. \eeq  

Spherical accretion flows produce x-rays by thermal bremsstrahlung and
by Comptonizing synchrotron photons.  We are interested in very low
luminosity systems where the former is expected to dominate.  Thermal
bremsstrahlung emission can also be expressed using a reaction
coefficient.  The number of x-rays of frequency $\nu$ produced per
second and per cm$^3$ is given by $n^2 R_\nu(\theta_e)$, where
$R_\nu(\theta_e) \approx \beta \theta_e^{-1/2} \exp[-h \nu/k T_e]$ and
$\beta \approx 1.3 \times 10^{-16}$ cm$^3$ s$^{-1}$ (e.g., Rybicki \&
Lightman 1979).  Integrating over the flow, the photon luminosity in
x-rays of frequency $\nu$ is \beq N_X = 4 \pi R_S^3 n_0^2 \beta
\int_{1}^{\infty} \left({dr \over r}\right) r^{3 - 2a} \theta^{-1/2}_e
\exp[-h\nu/k T_e]. \label{nxtemp} \eeq Equation (\ref{nxtemp}) shows
that at a frequency $\nu$ the x-ray emission is dominated by the
largest radius which satisfies $kT_e \gsim h \nu$.  This is because
$r^3 n^2 T_e^{-1/2}$ increases with increasing radius.  We focus on
x-ray emission at $\sim 10$ keV which is dominated by emission from $r
\sim 10^3-10^4$.  At these radii the flow is quite accurately
approximated as one-temperature so we can substitute $T_e = T_p$
($\theta_e = m_p \theta_p/m_e$) into equation (\ref{nxtemp}) and
perform the integral \beq N_X \approx {4 \pi \over 3.5 - 2a} R^3_S
n^2_0 \beta' \theta_0^{-1/2} r^{3.5-2a}_\nu,
\label{nx} \eeq where $r_\nu = \theta_0/\theta_\nu$,
$\theta_\nu = h\nu/m_p c^2$, and $\beta' = \beta (m_e/m_p)^{1/2}
\approx 3 \times 10^{-18}$ cm$^{3}$ s$^{-1}$.

Combining equations (\ref{ngamma}) and (\ref{nx}), the ratio of the
gamma-ray luminosity at energy $E_\gamma \approx 100$ MeV to the x-ray
luminosity at energy $E_X$ is given by \beq {L_\gamma \over L_X}
\approx \left({E_\gamma \over E_X}\right) \left(3.5 - 2a \over
a\right) r_\nu^{2a-3.5} \approx 30 \left( {10 \ {\rm keV}\over
E_X}\right) ^{1/2}, \label{ratio} \eeq where the last approximation
takes $a = 3/2$.

Equation (\ref{ratio}) shows that, for the self-similar analysis of
this subsection, the gamma-ray to x-ray luminosity ratio of the flow
depends only on the radial density profile.  Since both pion decay and
bremsstrahlung involve two-body processes the luminosity ratio from
any spherical shell depends only on the local temperature(s).  The
radial density profile enters because pions are only produced in
interesting numbers very close to the black hole while the x-ray
luminosity primarily originates from rather large radii.  For $a =
3/2$, equation (\ref{ratio}) predicts $L_\gamma \approx 30 L_X$, an
observationally interesting number (\S3), while for $a = 1/2$, the
predicted gamma-ray luminosity is certainly undetectable, $L_\gamma
\approx 10^{-5} L_X$.

\subsection{Uncertainties in Bondi-like models ($a = 3/2$)}

Several models of quasi-spherical accretion (Bondi, ADAF) predict a
nearly free-fall radial velocity and, as a consequence, $a = 3/2$. In
this case, the primary uncertainty in the $L_\gamma/L_X$ estimate of
the previous section is the proton temperature: $\theta_p$ could be
smaller than $\sim 0.1$ near the black hole. {\it A priori} this is
quite worrying because of the strong temperature dependence of the
pion reaction rate.

We do not believe that this uncertainty poses a serious threat to the
estimate of equation (\ref{ratio}). Relativistic corrections to the
classical Bondi solution are small (Shapiro 1973). As we explain
below, the corrections due to rotation in the ADAF solution are
relatively small as well.

In principle, ADAF models can have low proton temperatures near the
event horizon (e.g., $\theta_p \lsim 0.03$). The low-temperature
solutions, however, require small values of the dimensionless
viscosity $\alpha$, while numerical simulations and theoretical
arguments (see \S 2.2) show that canonical ADAF models are only
realizable if $\alpha$ is relatively large, roughly $\alpha \gsim
0.1$.

For large $\alpha$ equation (\ref{td}) is a reasonable approximation
of even general relativistic calculations of the structure of ADAFs
(Gammie \& Popham 1998, Popham \& Gammie 1998; hereafter
GP).\footnote{This is not true for small $\alpha$ (e.g., Narayan,
Kato, \& Honma 1997).}  For non-rotating black holes and $\alpha \gsim
0.1$, for example, our temperature profiles match those of GP very
well;\footnote{GP consider several adiabatic indices for the flow; we
compare only with $\gamma \approx 5/3$, appropriate for a flow
dominated by the energy density of the nearly non-relativistic
protons.}  they find maximal temperatures of $\theta_p \approx 0.1$,
consistent with our value.  For rapidly spinning black holes, their
temperatures are yet higher, reaching $\theta_p \approx 0.3$.  In
addition, the density profile given by equation (\ref{td}) is a
reasonable approximation of the global calculations for large
$\alpha$.  Self-similar solutions predict radial velocities $\sim
\alpha c_s$, where $c_s$ is the sound speed of the gas.  At small
radii, however, the accreting gas must pass through a sonic point on
its way into the black hole.  For large $\alpha$ the ``natural''
radial velocity of the flow is of order the sound speed, so little
deviation from self-similarity is required to match onto the sonic
transition.

To test the estimate of equation (\ref{ratio}) we calculated the
expected gamma-ray to x-ray luminosity ratio using several of GP's
models and found generally good agreement.  For a non-rotating black
hole and an accretion flow with $\alpha \approx 0.3$, for example, the
more detailed calculation yields $L_\gamma \approx 10 L_X$ for $E_X =
10$ keV, in reasonable agreement with equation (\ref{ratio}).

It is also important to emphasize that gamma-ray emission from pion
decay is unlikely to be as sensitive to temperature as suggested by
the simple thermal model we have considered.  The collisionless
plasmas of interest should efficiently accelerate protons to
relativistic energies; in this case the total gamma-ray luminosity
varies only linearly with changes in the thermal energy of the protons
since there are always a substantial number of protons above the pion
production threshold (see Mahadevan et al. 1997).\footnote{Gruzinov \&
Quataert (1999) describe a proton heating model which yields very
little proton acceleration. However, if shocks or reconnection events
occur in the accretion flow, a fraction of protons should be
accelerated.}

Equation (\ref{ratio}) predicts a detectable gamma-ray flux only if $a
\approx 3/2$.  The above considerations suggest that equation
(\ref{ratio}) should be a good approximation in this limit.

\subsection{Non Bondi-like accretion models ($a < 3/2$)}

Modern theories and numerical simulations of quasi-spherical accretion
flows suggest that the mean infall velocity can deviate substantially
from the free-fall value, resulting in $a<3/2$. There are three
scenarios: convection-dominated accretion flows (CDAFs), winds, and
turbulent heat conduction.

{\it CDAF:} Two independent groups have performed numerical
simulations of quasi-spherical non-radiating accretion flows with
small values of the viscosity parameter (Stone, Pringle, \& Begelman
1999; Igumenshchev \& Abramowicz 1999).  They both find $a = 1/2$
rather than $a = 3/2$.  Narayan, Igumenshchev, \& Abramowicz (2000)
and Quataert \& Gruzinov (2000) have explained this in terms of a
CDAF. In such a flow angular momentum is efficiently transported
inwards by strong radial convection.  This nearly cancels the outward
transport by magnetic fields, leading to a substantially suppressed
accretion rate and a much flatter radial density profile.

{\it Winds:} For large $\alpha \gsim 0.1$, CDAFs do not appear to be found in
numerical simulations (Igumenshchev \& Abramowicz 1999); this is
because the infall time of the gas is shorter than the convective
turnover time, so convection is less dynamically important.  For large
$\alpha$, however, $a$ may differ from $3/2$ for a different physical
reason; strong outflows may drive away most of the accreting mass
(Blandford \& Begelman 1999; Igumenshchev \& Abramowicz 1999).

{\it Turbulent heat conduction:} Conduction preheats the infalling gas, 
reducing the accretion rate and flattening the density profile
(Gruzinov 1999).

\section{Application to the Galactic Center}

Chandra observations of the Galactic Center detect a point source
coincident with the non-thermal radio source Sgr A* to within $\approx
0.5'' \approx 10^5 R_S$.  Its $0.1-10$ keV luminosity is $L_X \approx
4 \times 10^{33}$ ergs s$^{-1}$ (Baganoff et al. 2000).  It is very
plausible that this represents the first x-ray detection of the
supermassive black hole at the center of our galaxy.

It is natural to interpret Chandra's detection as thermal
bremsstrahlung from large radii in the accretion flow.  As shown in
\S2, such emission would arise from $r \sim 10^4$; the density
required to match the observed luminosity is then $\approx 4 \times
10^3$ cm$^{-3}$.  The corresponding accretion rate is $\approx
10^{-5}$ $M_\odot$ yr$^{-1}$, if the radial velocity of the gas is of
order the sound speed.  This is in good agreement with estimates based
on the mass-losing stars in the central parsec of the Galactic Center
(e.g., Coker \& Melia 1997; Quataert, Narayan, \& Reid 1999).  The
bremsstrahlung interpretation predicts the absence of short timescale
variability in the observed x-rays (since the emission arises from
large radii).  It can also be tested by looking for x-ray line
emission in deeper Chandra exposures (Narayan \& Raymond 1999).

EGRET observations of the Galactic Center region detect a source (2EG
1746-2852) with $L_\gamma \approx 10^{36}$ ergs s$^{-1}$ and a power
law spectrum extending from $\approx 100$ MeV to $\approx 10$ GeV
(Merck et al. 1996); it appears to be point like within the $\approx
1^o$ resolution of the instrument.  Mahadevan et al. (1997)
interpreted this emission as arising from an ADAF around the black
hole at the Galactic Center.  In their more comprehensive models of
Sgr A*, however, Narayan et al. (1998) were unable to produce
gamma-ray emission at the required levels and satisfy other
observational constraints.  Moreover, the observed spectrum of 2EG
1746-2852 looks very similar to that expected from cosmic rays
colliding with a dense cloud of molecular hydrogen.

Our calculation in \S2 predicts the expected gamma-ray luminosity from
the accretion flow given the x-ray luminosity in thermal
bremsstrahlung.  We believe that the Chandra observations of the
Galactic Center provide this thermal bremsstrahlung luminosity.  With
this normalization, equation (\ref{ratio}) predicts $L_\gamma \approx
10^{35}$ ergs s$^{-1}$ for $a = 3/2$.  This is well above the $\approx
2 \times 10^{33}$ ergs s$^{-1}$ detection threshold of the forthcoming
Gamma-ray Large Area Space Telescope (GLAST).\footnote{see
http://glast.gsfc.nasa.gov/SRD} In addition, GLAST's angular
resolution is expected to be significantly better than that of EGRET
(for, among other things, the express purpose of identifying
unidentified EGRET sources).  GLAST will therefore likely have the
capability of distinguishing a gamma-ray counterpart of Sgr A* (if it
indeed exists) from 2EG 1746-2852.


\section{Discussion}

We have argued that the ratio of the $\approx 100$ MeV gamma-ray
luminosity to the $\sim 10$ keV x-ray luminosity of a spherical
accretion flow depends primarily on its radial density profile (where
$n \propto r^{-a}$).  In particular, for canonical spherical accretion
flow models with $a = 3/2$, $L_\gamma \approx 30 L_X$; for $a < 3/2$,
$L_\gamma \ll L_X$.

Our analysis predicts the $\approx 100$ MeV gamma-ray luminosity
expected from the accretion flow onto the supermassive black hole at
the center of our galaxy (\S3): $L_\gamma \approx 10^{35}$ ergs
s$^{-1}$ if $a = 3/2$ while $L_\gamma \approx 10^{29}$ ergs s$^{-1}$
if $a = 1/2$.  For $a = 3/2$, this estimate is nearly two orders of magnitude
above the detection threshold of the GLAST telescope.  We expect,
however, that no gamma-rays will be observed coincident with the black
hole, supporting theoretical suggestions (see \S2.2) that the density
profile in spherical accretion flows is significantly flatter than the
canonical $r^{-3/2}$ profile.


\acknowledgments We thank Charles Gammie and Bob Popham for the use of
their global models of ADAFs, Fred Baganoff for information about
Chandra observations of the Galactic Center, and John Bahcall for
discussions and comments.  EQ is supported by NASA through Chandra
Fellowship PF9-10008, awarded by the Chandra X--ray Center, which is
operated by the Smithsonian Astrophysical Observatory for NASA under
contract NAS 8-39073. AG was supported by the W. M. Keck Foundation
and NSF PHY-9513835.

\end{document}